\documentclass[
 reprint,
 superscriptaddress,
 amsmath,amssymb,
 aps,
pra,
showkeys
]{revtex4-2}

\usepackage{soul}
\usepackage{float}
\usepackage{multibib}
\usepackage{hyperref}
\usepackage{xcolor}
\hypersetup{
    colorlinks,
    linkcolor={blue!50!blue},
    citecolor={blue!50!blue},
    urlcolor={blue!80!black}
}
\usepackage{siunitx}
\usepackage{import}
\usepackage{placeins}
\usepackage{xcolor}
\usepackage{lmodern}
\usepackage[braket, qm]{qcircuit}
\usepackage{graphicx}
\usepackage{dcolumn}
\usepackage{bm}
\usepackage{comment}
\usepackage[english]{babel}
\usepackage{blindtext}
\usepackage{physics}  
\usepackage{multirow}
\usepackage{array}
\usepackage{enumerate}

\usepackage{ulem}

\begin{document}
\title{Quantum Emitters at Telecommunication Wavelengths based on Carbon Defects in Transition Metal Dichalcogenides}  

\author{Chanaprom Cholsuk}
\email{chanaprom.cholsuk@tum.de}
\affiliation{Department of Computer Engineering, TUM School of Computation, Information and Technology, Technical University of Munich, 80333 Munich, Germany}
\affiliation{Munich Center for Quantum Science and Technology (MCQST), 80799 Munich, Germany}

\author{Sujin Suwanna}
\affiliation{Optical and Quantum Physics Laboratory, Department of Physics, Faculty of Science, Mahidol University, Bangkok 10400, Thailand}

\author{Tobias Vogl}%
\email{tobias.vogl@tum.de}
\affiliation{Department of Computer Engineering, TUM School of Computation, Information and Technology, Technical University of Munich, 80333 Munich, Germany}
\affiliation{Munich Center for Quantum Science and Technology (MCQST), 80799 Munich, Germany}

\date{\today}

\begin{abstract}
Low-dimensional materials have emerged as promising hosts for quantum emitters, whose emission typically arises from either strain-induced band bending or defect-induced two-level systems. Among these materials, transition metal dichalcogenide (TMD) monolayers have attracted particular attention; however, their performance is limited by strong photoluminescence (PL) quenching at room temperature. As TMDs transition from a direct to an indirect bandgap when moving from monolayers to multilayers, we herein propose a strategy to overcome this quenching limitation by exploiting the indirect bandgap of TMD bilayers in combination with a point defect doping. The indirect gap suppresses excitonic PL, while specific defects enable robust defect-mediated quantum emission. Using hybrid-functional density functional theory, we investigate substitutional carbon defects at chalcogen sites (S and Se) in WS$_2$, WSe$_2$, MoS$_2$, and MoSe$_2$ bilayers and comprehensively characterize their optical properties. Both neutral and singly negative charge states are found to be thermodynamically stable. Neutral defects exhibit singlet configurations with emission in the O- and C-band telecommunication windows, whereas negatively charged defects adopt doublet configurations featuring spin-selective transitions and near-infrared emission. The electron–phonon coupling strength, radiative lifetime, and dipole orientation are found to depend sensitively on both the host material and defect site, providing distinct fingerprints for experimental identification. Our findings, therefore, establish carbon-doped TMD bilayers as promising platforms for room-temperature defect-based quantum emitters operating at telecommunication wavelengths.
\end{abstract}


\maketitle

\section{Introduction}
Atomically thin van der Waals materials have emerged as promising platforms for solid-state quantum emitters (QEs), complementing well-established systems such as diamond color centers \cite{10.1016/j.physrep.2013.02.001,10.1126/science.1139831} and semiconductor quantum dots \cite{10.1038/nnano.2017.218,10.1515/nanoph-2019-0007}. Their planar geometry and inherently low refractive indices can enable high photon out-coupling efficiencies, making them attractive for integrated quantum photonic applications.~Among these materials, hexagonal boron nitride (hBN) \cite{10.1038/nnano.2015.242,10.1063/5.0147560,10.1021/acsnano.3c08940} and transition-metal dichalcogenides (TMDs) \cite{10.1038/s41467-019-10632-z,10.1038/nnano.2013.277,10.3390/nano13091501,10.1038/ncomms15093,10.1038/ncomms15093} have attracted particularly high attention.\\
\indent Defects in hBN exhibit several favorable optical characteristics, including strong compatibility with photonic architectures \cite{10.1088/1361-6463/aa7839,10.1002/qute.2023003433}, exceptional photostability \cite{10.1021/acsphotonics.8b00127,10.1016/j.physe.2020.114251}, bright and spectrally pure single-photon emission \cite{10.1021/acsnano.3c08940}, and a wide variety of possible defect formation \cite{10.3390/nano12142427,10.1021/acs.jpcc.4c03404,10.1039/D5TC02805A}. Nevertheless, hBN-based QEs suffer from significant decoherence compared to the diamond color centers \cite{10.1038/s41524-025-01859-0, 10.1126/science.1220513}, which limits their spin-related properties and hence hinders nuclear-spin-based quantum applications \cite{10.1002/adom.202402508,10.1002/adom.202302760}.\\
\indent In contrast, TMDs offer the advantage of reduced decoherence compared to hBN, leading to longer coherence times \cite{10.1038/s41524-019-0182-3,10.1103/PhysRevB.106.104108} while they are also attractive for photonic integration as they can be easily grown in arbitrary patterns \cite{10.1002/smtd.202200300}. However, their main limitation lies in the lack of stable and efficient single-photon emission, especially at room temperature. 
For TMD monolayers, the emitters were first discovered in WSe$_2$ \cite{10.1364/OPTICA.2.000347,10.1038/ncomms15093,10.1038/s41699-020-0136-0,10.1038/s41467-021-23709-5,10.1038/nnano.2015.75} and subsequently in a variety of TMDs, including MoS$_2$ \cite{10.1038/s41467-019-10632-z,10.1021/acsnano.2c09209}, MoSe$_2$ \cite{10.1038/nnano.2013.277,10.1021/acs.nanolett.0c04282}, WS$_2$ \cite{10.1038/ncomms15093,10.1038/s41467-021-27585-x,10.1126/sciadv.abb5988}, and MoTe$_2$ \cite{10.3390/nano13091501,arxiv:2508.20743}. However, the microscopic origin of single-photon emission remains under debate. The most likely mechanisms involve exciton localization induced either by strain-induced band bending, defect-induced intervalley-bound states, or a combination of both. In all cases, the emission is associated with excitonic transitions between the conduction and valence bands.\\
\indent Experimentally, these quantum emitters are mostly observed at cryogenic temperatures, since the localized (0D) excitons that enable sharp single-photon emission are confined by potential wells with depths on the order of tens to a few hundred meV \cite{10.1021/acsnano.9b02316,10.1021/acs.nanolett.5b03312,10.1038/s41467-021-23709-5}. At low temperatures, thermal energy is insufficient to overcome the localization potential, allowing stable single-photon emission. As the temperature increases, however, thermal energy ($k_BT$) becomes comparable to or larger than the confinement energy, enabling excitons to delocalize or escape to other band states. Even if an exciton remains localized at elevated temperatures, strong exciton–phonon scattering induces thermal broadening of the zero-phonon line (ZPL), thereby degrading single-photon purity. In addition, exciton–phonon coupling enhances nonradiative recombination. Together, these processes lead to quenching of 0D exciton emission at room temperature. Despite significant progress, WSe$_2$ monolayer QEs currently preserve their single-photon nature only up to $\sim$160 K \cite{10.1038/s41467-021-23709-5,10.1088/2053-1583/ab15fe}, leaving robust room-temperature operation as a key open challenge.\\
\indent Another important consideration is the role of layer stacking in TMDs. While monolayers possess a direct bandgap at the $K$ and $K'$ valleys \cite{10.1103/PhysRevLett.105.136805}, bilayers and multilayers typically exhibit an indirect bandgap \cite{10.1103/PhysRevLett.105.136805,10.1021/nl302584w,10.1021/nn305275h,10.1021/nn305275h}. As such, in this work, we propose to exploit this indirect-gap property as a pathway to overcome the temperature limitations of monolayer QEs. In monolayers, direct-gap transitions facilitate radiative recombination of both free excitons and localized excitons, contributing to a strong photoluminescence (PL) background at low temperature but quenching at elevated temperatures. In contrast, the indirect bandgap in bilayers suppresses radiative recombination of free excitons, thereby reducing the background emission and opening the door for alternative quantum emission mechanisms. In our approach, we do not rely on localized excitons formed through strain-induced band bending; instead, we introduce point defects to create discrete in-gap electronic states that form a two-level system. Single-photon emission then arises from transitions between defect states, rather than from excitonic recombination. Since defect states can reside deep within the bandgap, such emitters have the potential to remain stable even at room temperature.\\
\indent As a consequence, we investigate carbon substitutional doping at chalcogen sites in WS$_2$, WSe$_2$, MoS$_2$, and MoSe$_2$ bilayers using density functional theory (DFT) based on Heyd–Scuseria–Ernzerhof (HSE06) functional \cite{10.1063/1.1564060}. While other quantum emitters based on hBN and NV color centers have wide band gap, which leads to the emission wavelength mostly in the visible, these TMD bilayers have the smaller bandgap of approximately 1.75 eV for WS$_2$ \cite{10.1021/nn305275h}, 1.6 eV for WSe$_2$ \cite{10.1021/nn305275h}, 1.6 eV for MoS$_2$ \cite{10.1103/PhysRevLett.105.136805}, and 1.3 eV for MoSe$_2$ \cite{10.1021/nl302584w}.  We therefore anticipate that the defect states are formed with transition energies close to the telecommunication band. This, therefore, can give us the telecomm ZPL. In this work, we also characterize key photophysical properties, including the spin configurations, charge stability, ZPL, Huang–Rhys (HR) factor, Debye–Waller (DW) factor, PL spectrum, transition rates, lifetimes, and dipole orientations. Our work thus moves beyond the widely studied monolayer localized-exciton QEs, highlighting bilayer defect engineering as a promising and potentially more robust route toward room-temperature quantum light sources.

\begin{figure*}[ht!]
    \centering
    \includegraphics[width=1\linewidth]{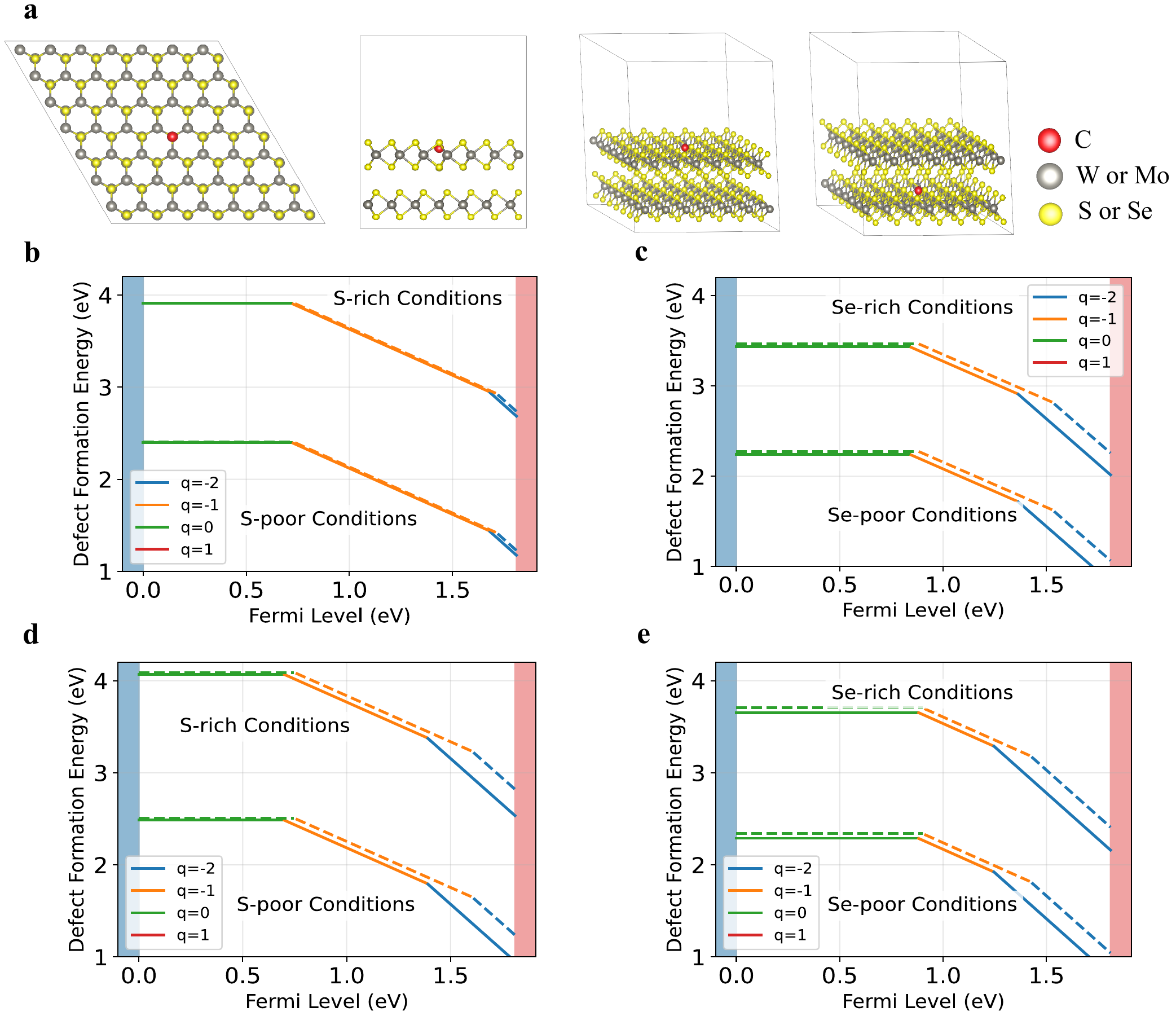}
    \caption{\textbf{Geometries and charge stabilities.} \textbf{a} The 7$\times$7$\times$1 supercell structures where the carbon is doped on top or middle layers. \textbf{b} - \textbf{e} Defect formation energies of WS$_2$, WSe$_2$, MoS$_2$, and MoSe$_2$, respectively, under chalcogen-rich and chalcogen-poor conditions as a function of Fermi level. Finite-size charge corrections were taken into account. The Fermi level represents the chemical potential, which varies across the bandgap. The crossing point between the formation energy lines of different charge states $(q)$ defines the charge transition level. Solid lines correspond to structures with C doped at the topmost layer, while dashed lines correspond to structures with C doped at the middle layer.}
    \label{fig:Eformation}
\end{figure*}

\section{Results}
The analyses of bilayer QEs are divided into four parts. Firstly, we investigate the ground-state properties of the four mentioned host materials, including their pristine electronic bandgaps, the preferred spin configurations of defect systems, and charge stability. Secondly, for defects in their most stable spin and charge states, we characterize the electronic transitions, where selected transition pathways are used to compute the ZPL values. Thirdly, we analyze the PL spectra in conjunction with electron–phonon coupling, quantified by the HR factor. Finally, we present additional optical properties, including radiative lifetimes and dipole orientations. Calculating this complete optical fingerprint allows us to judge quantitatively the usefulness of the emitter for quantum technology applications.

\subsection{Ground-state properties}
We begin by examining the bandgap of each pristine bilayer host. Using spin-polarized DFT with the HSE06 hybrid functional \cite{10.1063/1.1564060} (see Methods for further details), our calculations yield bandgaps of 1.81 eV, 1.74 eV, 1.68 eV, and 1.44 eV for WS$_2$, WSe$_2$, MoS$_2$, and MoSe$_2$, respectively. These values are in good agreement with the experimentally observed optical bandgaps \cite{10.1103/PhysRevLett.105.136805,10.1021/nl302584w,10.1021/nn305275h,10.1021/nn305275h}. Therefore, the HSE06 functional is employed throughout this study. For defect modeling, we construct a $7\times 7\times 1$ supercell and introduce carbon substitution at either a sulfur or selenium site, which will be denoted as C$_\text{S}$ and C$_\text{Se}$, respectively. This supercell size was optimized to mitigate the finite size effects, as can be seen in Supplementary Section S1. To capture environmental effects, we consider two nonequivalent substitutional sites: (i) the top chalcogen site located at the outermost surface of the bilayer and directly exposed to vacuum, and (ii) the middle chalcogen site located in the upper layer of the bottom monolayer and thus adjacent to the second monolayer above it (see the geometries in Fig.~\ref{fig:Eformation}\textbf{a}). These two configurations experience distinct local environments, which can influence both the defect energies and the resulting optical properties. Owing to the symmetry of the bilayer structure, all other substitutional sites are symmetry-equivalent and therefore yield identical results.  \\
\indent To assess charge stability, we compute the defect formation energies including correction terms (see Methods for computational details). The charge states were systematically varied until unstable configurations were identified. Accordingly, the charge states considered in this study are –2, –1, 0, and +1. As depicted in Fig.~\ref{fig:Eformation}, we found that the thermodynamically stable charge states are –2, –1, and 0. The charge transition level (CTL) for the $(0|-1)$ transition under both chalcogen-rich and chalcogen-poor conditions is located at approximately 0.75 eV, 0.90 eV, 0.79 eV, and 0.97 eV for WS$_2$, WSe$_2$, MoS$_2$, and MoSe$_2$, respectively. Note that chalcogen-rich and chalcogen-poor conditions refer to the thermodynamic limits of the experimental synthesis environment. We also found that the CTLs are nearly identical for the top- and middle-layer substitutions, indicating that the local environment has little influence on the defect energies. Although the doubly negative charge state exhibits thermodynamic stability, its CTL of some TMDs lies close to the band edge, making it highly improbable that transitions involving this state will physically occur. Furthermore, no experimental evidence for the –2 state has been reported to date. Consequently, we restrict our focus to the neutral and singly negative charge states as the preferred configurations across all host materials.

\begin{figure*}[ht!]
    \centering
    \includegraphics[width=1\linewidth]{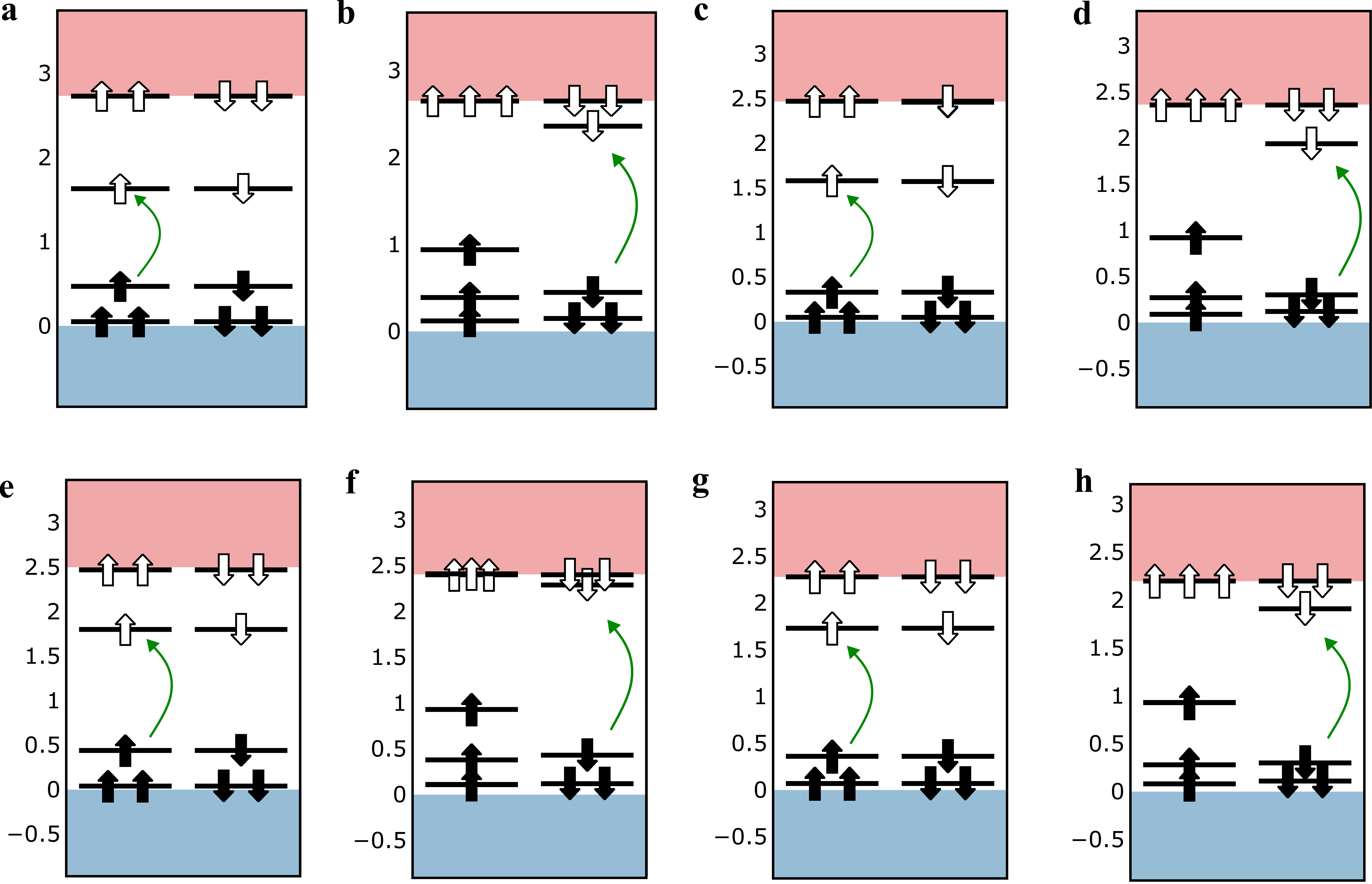}
    \caption{\textbf{Kohn–Sham electronic transitions at the $\Gamma$ point for carbon substitution in the top layer.} \textbf{a} and \textbf{b} for WS$_2$ doped with C$_\text{S}$ and C$_\text{S}^{-1}$. \textbf{c} and \textbf{d} for WSe$_2$ doped with C$_\text{Se}$ and C$_\text{Se}^{-1}$. \textbf{e} and \textbf{f} for MoS$_2$ doped with C$_\text{S}$ and C$_\text{S}^{-1}$. \textbf{g} and \textbf{h} for MoSe$_2$ doped with C$_\text{Se}$ and C$_\text{Se}^{-1}$. The green arrows represent the electronic transition responsible for the computed ZPLs. Note that the electronic transitions for the doping at the middle layer remain unchanged, as can be seen from Supplementary S2.}
    \label{fig:transition}
\end{figure*}

\begin{figure*}[ht!]
    \centering
    \includegraphics[width=0.9\linewidth]{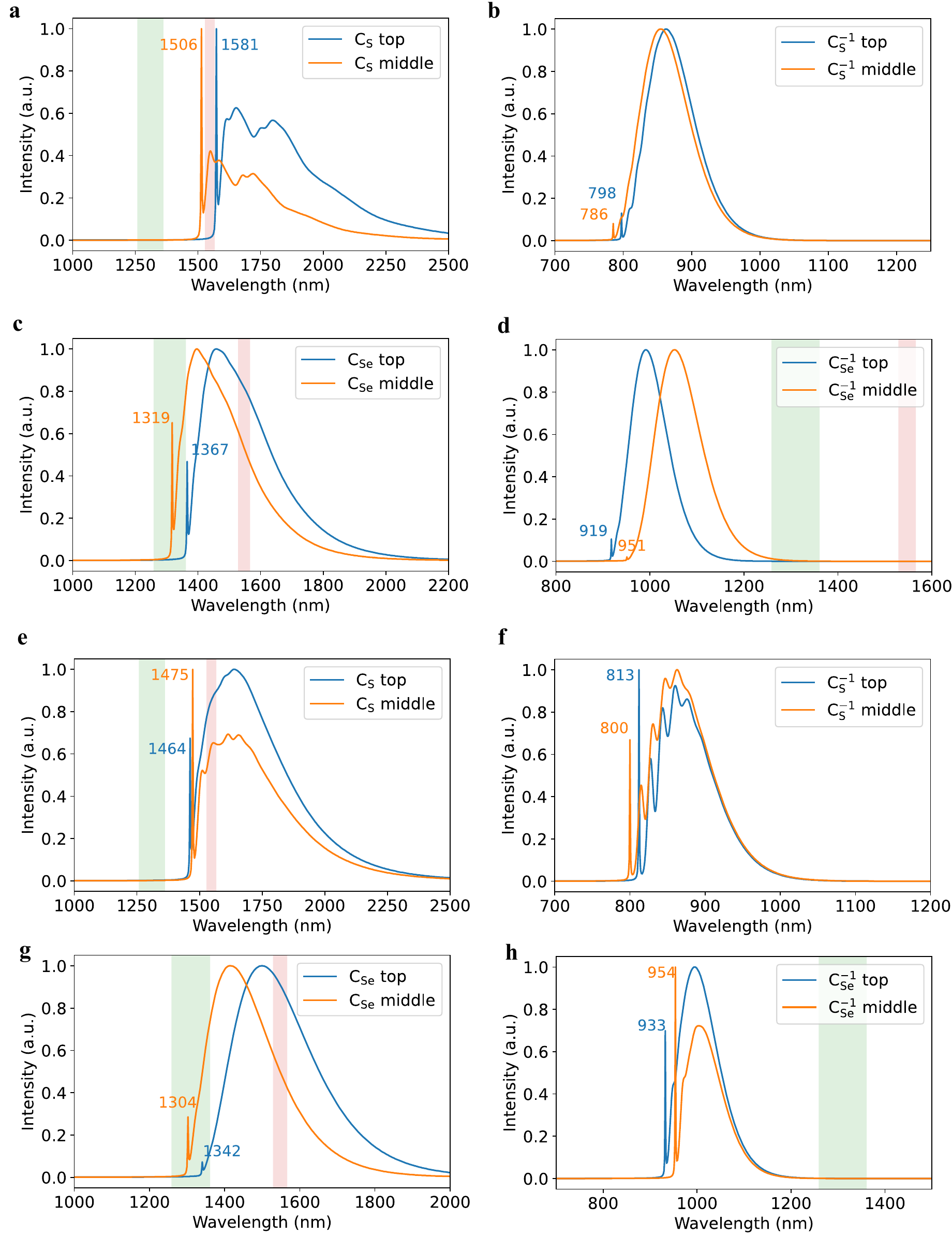}
    \caption{\textbf{Simulated photoluminescense spectra.} \textbf{a} and \textbf{b} for WS$_2$ doped with C$_\text{S}$ and C$_\text{S}^{-1}$. \textbf{c} and \textbf{d} for WSe$_2$ doped with C$_\text{Se}$ and C$_\text{Se}^{-1}$. \textbf{e} and \textbf{f} for MoS$_2$ doped with C$_\text{S}$ and C$_\text{S}^{-1}$. \textbf{g} and \textbf{h} for MoSe$_2$ doped with C$_\text{Se}$ and C$_\text{Se}^{-1}$. The green and pink shaded regions indicate the telecommunication wavelength ranges for the O-band (1260-1360 nm) and C-band (1530-1565 nm), respectively.}
    \label{fig:PL}
\end{figure*}

\subsection{Electronic transitions}
Next, we examine the preferred spin configurations and their corresponding ZPLs. As shown in Fig.~\ref{fig:transition}, we found that the carbon substituted C$_\text{S}$ and C$_\text{Se}$ defects in all four bilayer hosts favor the singlet configuration ($S=0$), whereas C$_\text{S}^{-1}$ and C$_\text{Se}^{-1}$ favor the doublet configuration ($S=1/2$).\\ 
\indent For neutral defects, which belong to the singlet manifold, we consider only spin-up to spin-up transitions when calculating the ZPL, as the spin-down pathway will yield the same ZPL value. To simulate the excited-state configuration, we employed the $\Delta$SCF method to manually promote one electron from the occupied defect state to the unoccupied defect state \cite{10.1103/RevModPhys.61.689}. The total energy difference between the ground and excited states was then used to evaluate the ZPL. The resulting ZPL energies are 0.78 eV, 0.91 eV, 0.85 eV, and 0.92 eV for the top-layer substitutions in WS$_2$, WSe$_2$, MoS$_2$, and MoSe$_2$, respectively. The corresponding middle-layer substitutions yield very similar ZPL values of 0.82 eV, 0.94 eV, 0.84 eV, and 0.95 eV. These values lie within the telecommunication windows, namely the O-band (0.91–0.98 eV) and the C-band (0.79–0.81 eV). The ZPL energies are, however, sensitive to finite-size effects. In this work, tests up to a $9 \times 9 \times 1$ supercell show a deviation of only 5 meV compared with the $7 \times 7 \times 1$ supercell used here. This small deviation confirms that the computed ZPL values are well converged. Therefore, ZPLs lying within a few meV of the O- or C-band boundaries can reasonably be considered consistent with these telecommunication windows. As shown in Fig.~\ref{fig:transition}, this telecom emission arises from the well-localized defect states of the neutral charge configurations within the bandgap, which give rise to optical transition energies in the telecom range.  It should be noted that although these ZPLs lie close to the CTLs, which could in principle induce charge ionization, the neutral charge states can remain thermodynamically stable. Experimentally, the neutral C$_\text{Se}$ defects in WSe$_2$ monolayers have been observed to be stable and adopt a singlet configuration \cite{10.1038/s41467-021-27585-x}. This stability arises from the band alignment between WSe$_2$ and the substrate (e.g., graphene and SiC), which positions the Fermi level near mid-gap and prevents electron transfer from the substrate into the defect state. Thus, the C$_\text{Se}$ defect become the neutral charge \cite{10.1038/s41467-021-27585-x}.\\
\indent For negatively charged defects (C$_\text{S}^{-1}$ and C$_\text{Se}^{-1}$), which adopt a doublet configuration, ZPLs were computed for the spin-down to spin-down transition, as illustrated in Fig.~\ref{fig:transition}. The corresponding ZPLs are 1.55 eV, 1.35 eV, 1.53 eV, and 1.33 eV for the top-layer substitutions, and 1.58 eV, 1.30 eV, 1.55 eV, and 1.30 eV for the middle-layer substitutions. As indicated in Fig.~\ref{fig:transition}, spin-up transitions are less favorable because the unoccupied defect states involved lie close to the conduction band edge, making them prone to thermal activation. In contrast, the position of spin-down transitions is more ideal and therefore expected to dominate the optical response. However, unlike the neutral case, the ZPLs of singly negative defects fall outside the telecom band due to the wider two-level defect separation, but remain within the near-infrared range. When taking the charge stability into account, we note that while C$_\text{Se}$ defects in WSe$_2$ monolayers have been experimentally observed in the neutral charge state with a singlet configuration, C$_\text{S}$ defects in WS$_2$ monolayers have instead been reported to stabilize in the negative charge state C$_\text{S}^{-1}$, and favor a doublet configuration \cite{10.1038/s41467-021-27585-x}, consistent to our finding.\\
\indent Overall, both the neutral and singly negative charge states are thermodynamically accessible, and their relative stability is likely to depend on the specific host environment. Our results for charge and spin configurations are in agreement with existing experimental observations in monolayers \cite{10.1038/s41467-021-27585-x}. However, since these experimental studies have not yet explored bilayer systems, systematic experimental investigations in bilayers are still required. To provide a complete theoretical picture, we next analyze the PL spectra and associated optical properties for both neutral and negatively charged defects.

\subsection{Photoluminescense spectra}
Having identified the ZPL energies, we next evaluate the full PL spectra of all considered defects. These spectra allow us to determine the origin of the telecom-relevant emission, whether from the ZPL, the PSB, or both, and to quantify the strength of electron--phonon coupling through the HR and DW factors. These quantities are critical for assessing the viability of these defects for quantum applications. The phonon modes for each defect are calculated over all $3N$ degrees of freedom, where $N$ is the total number of atoms, using the Pyphotonics Python package \cite{10.1016/j.cpc.2021.108222}. Additional computational details are provided in Sec.~\ref{sec:PL}.\\
\indent For neutral defects, Figs.~\ref{fig:PL}\textbf{a}, \textbf{c}, \textbf{e}, and \textbf{g} show that all four TMD bilayers exhibit emission within or near the telecommunication O- and C-band windows, arising from both ZPL and PSB contributions. In contrast, for the negatively charged defects, Figs.~\ref{fig:PL}\textbf{b}, \textbf{d}, \textbf{f}, and \textbf{h} suggest that the ZPLs lie in the near-infrared range, as discussed in the previous section.\\
\indent In WS$_2$, as illustrated in Fig.~\ref{fig:PL}\textbf{a}, the neutral C$_\text{S}$ defects in the top and middle layers yield similar ZPLs, but slightly different PSB strengths. The ZPLs lie close to the C-band telecommunication window, indicating that the main emission in the C-band range is from the ZPLs of this defect. The HR factor is 3.34 for the top layer and 2.77 for the middle layer, corresponding to the configuration coordinate $Q$ of 1.24 and 1.10, respectively. Larger $Q$ values indicate stronger lattice relaxation difference between ground and excited states, and thus stronger PSB contributions. For the C$_\text{S}^{-1}$ defects, as displayed in Fig.~\ref{fig:PL}\textbf{b}, both sites turn to produce similar results in ZPL, which is around 1.55 eV and 1.58 eV, and HR factors whose values are 5.05 and 5.50, consistent with structural change differences $Q=1.42$ and $1.51$ for the top and middle layers, respectively.\\
\indent In WSe$_2$, as shown in Fig.~\ref{fig:PL}\textbf{c}, neutral C$_\text{Se}$ defects at the top and middle layers exhibit comparable ZPLs and HR factors. The ZPLs are 0.91 and 0.94 eV for the top- and middle-layer defects, respectively, placing both within the O-band telecommunication window. Their HR factors are 4.17 and 3.83, consistent with the corresponding configuration coordinates of $Q=1.50$ and $Q=1.41$. In addition, the associated PSBs extend toward the C-band region, suggesting that this defect can contribute to emission across both O- and C-band wavelengths. For C$_\text{Se}^{-1}$ defects, shown in Fig.~\ref{fig:PL}\textbf{d}, a similar site dependence is observed. The ZPLs differ by only 0.05 eV, while the HR factors are 5.13 and 6.81 for the top- and middle-layer defects, respectively, with corresponding $Q$ values of 1.58 and 1.96.\\
\indent In MoS$_2$, as depicted in Fig.~\ref{fig:PL}\textbf{e}, neutral C$_\text{S}$ substitutions at the top and middle layers yield similar spectra, with HR factors of 4.07 and 3.39, respectively, corresponding to $Q=1.38$ and $Q=1.14$. Although the ZPLs are slightly shifted away from both the O- and C-band windows, their PSBs extend into the C-band region. Thus, this defect can still contribute to C-band emission through its phonon sideband. In contrast, C$_\text{S}^{-1}$ defects, as shown in Fig.~\ref{fig:PL}\textbf{f}, show nearly identical ZPLs for the two sites. The top-layer defect exhibits a pronounced PSB with HR = 3.28 and $Q=0.99$, while the middle-layer defect produces a slightly stronger PSB with HR = 3.76 and $Q=1.07$.\\
\indent In MoSe$_2$, as depicted in Fig.~\ref{fig:PL}\textbf{g}, neutral C$_\text{Se}$ defects exhibit a site-dependent ZPL shift of about 40 nm, while their PSB strengths differ more noticeably. The HR factors are 5.94 and 4.52 for the top- and middle-layer defects, respectively, with corresponding $Q=1.94$ and $Q=1.57$. For this defect, the ZPLs fall within the O-band telecommunication window, while the PSBs extend into the C-band region. For C$_\text{Se}^{-1}$ defects, shown in Fig.~\ref{fig:PL}\textbf{h}, the ZPLs are similar, at approximately 1.33 and 1.30 eV for the top and middle layers, respectively. However, the HR factors again show site dependence, with values of 3.59 for the top layer and 2.93 for the middle layer, consistent with the corresponding $Q$ values of 1.21 and 1.06.\\
\indent Taken together, these results show that the defect site, i.e., top versus middle layer, has only a minor influence on the ZPL energies and electron--phonon coupling strength for both charge states. For neutral defects, telecom-relevant emission can arise from the ZPL, the PSB, or both, depending on the host material. In contrast, negatively charged defects exhibit ZPLs in the near-infrared range. The remaining site-dependent variations mainly originate from differences in structural relaxation upon excitation, which are reflected in the HR factors and configuration coordinates. These subtle but systematic differences may therefore provide experimental fingerprints for distinguishing defect locations.

\begin{table*}[ht!]
\caption{Summary of the optical properties of carbon defects for both neutral and singly negative charge states in bilayer TMDs. $\theta$ is the angle relative to the $z$-axis, and $\phi$ is the angle relative to the $x$-axis in the $xy$-plane. $ex$ and $em$ denote excitation and emission, respectively. Misalignment refers to the angular difference between the excitation and emission dipoles.}
\resizebox{\textwidth}{!}{
\begin{tabular}{|c|c c c c c c | c c | c c | c|}
\hline Defects & Q & ZPL (eV) & ZPL (nm) & HR & DW factor & Lifetime (ns) & $\theta_{ex}$ ($^\circ$) & $\phi_{ex}$ ($^\circ$)& $\theta_{em}$ ($^\circ$)& $\phi_{em}$ ($^\circ$)& Misalignment ($^\circ$)\\
\hline WS$_2$-C$_\text{S}$-top & 1.24& 0.78& 1581& 3.34& 0.04& 87.86& 0.01& 26.83& 0.00 & 14.09& 0.01\\
WS$_2$-C$_\text{S}$-middle & 1.10& 0.82& 1506& 2.77& 0.06& 451.73& 0.00 & 51.37& 0.01& 26.19& 0.00 \\
\hline WSe$_2$-C$_\text{Se}$-top & 1.50& 0.91& 1367& 4.17& 0.02& 146.84& 1.56& 20.54& 0.01& 25.24& 1.56\\
WSe$_2$-C$_\text{Se}$-middle & 1.41& 0.94& 1319& 3.83& 0.02& 303.64& 0.00 & 66.61& 0.05 & 13.29& 0.05 \\
\hline MoS$_2$-C$_\text{S}$-top & 1.38& 0.85& 1464& 4.07& 0.02& 96.27& 0.16& 80.79& 0.05& 39.32& 0.12\\
MoS$_2$-C$_\text{S}$-middle & 1.14& 0.84& 1475& 3.39& 0.03& 452.31& 0.01& 74.38& 0.04& 29.77& 0.03\\
\hline MoSe$_2$-C$_\text{Se}$-top & 1.94& 0.92& 1342& 5.94& 0.00& 153.65& 1.22& 2.81& 0.00 & 63.39& 1.22\\
MoSe$_2$-C$_\text{Se}$-middle & 1.57& 0.95& 1304& 4.52& 0.01& 336.46& 0.00 & 35.05& 0.00 & 38.83& 0.00 \\
\hline WS$_2$-C$_\text{S}^{-1}$-top  & 1.42& 1.55& 798& 5.05& 0.01 & 293.40& 0.02& 79.21& 0.01& 26.70& 0.02\\
WS$_2$-C$_\text{S}^{-1}$-middle & 1.51& 1.58 & 786& 5.50& 0.00& 387.09& 0.00& 60.72& 0.02& 25.21& 0.02\\
\hline WSe$_2$-C$_\text{Se}^{-1}$-top & 1.58& 1.35& 919& 5.13& 0.01& 346.63& 0.40& 30.87& 0.01& 73.02& 0.39\\
WSe$_2$-C$_\text{Se}^{-1}$-middle & 1.96& 1.30& 951& 6.81& 0.00 & 213.32& 0.47& 60.16& 0.20& 63.59& 0.27\\
\hline MoS$_2$-C$_\text{S}^{-1}$-top & 0.99& 1.53& 813& 3.28& 0.04& 432.62& 0.04& 24.57& 0.06& 13.15& 0.02\\
MoS$_2$-C$_\text{S}^{-1}$-middle & 1.07& 1.55 & 800& 3.76& 0.02& 325.70& 0.00& 39.15& 0.01& 58.49& 0.01\\
\hline MoSe$_2$-C$_\text{Se}^{-1}$-top & 1.21& 1.33& 933& 3.59& 0.03& 493.62& 0.14& 3.08& 0.01& 11.25& 0.13\\
MoSe$_2$-C$_\text{Se}^{-1}$-middle & 1.06& 1.30 & 954& 2.93& 0.05& 212.78& 0.25& 46.79& 0.65& 42.44& 0.41\\
\hline
\end{tabular}
}
\label{tab:property}
\end{table*}

\subsection{Optical properties}
Finally, we characterize the radiative lifetimes and dipole orientations of the defects. Because these optical properties are uniquely determined by the specific atomic and electronic structure of each system, they serve as essential signatures for experimental defect identification. As summarized in Tab.~\ref{tab:property}, the lifetimes vary significantly with both the host material and defect position, ranging from tens to a few hundred nanoseconds. For both excitation and emission, the dipoles are nearly aligned along the out-of-plane $z$-axis for all considered systems, with only minor angular deviations. In particular, the excitation and emission angles relative to the $z$-axis remain below $1.56^\circ$ and $0.65^\circ$, respectively, indicating negligible in-plane dipole components. The small excitation-emission misalignment further suggests that the optical transitions preserve nearly the same dipole orientation. These distinct optical characteristics highlight the sensitivity of defect behavior to both the host material and doping configuration, and may therefore serve as useful fingerprints for experimental identification.

\section{Discussion}
In this work, we have proposed a strategy to overcome the PL quenching that typically limits monolayer TMD quantum emitters at elevated temperatures. By exploiting the indirect bandgaps of bilayers, the PL from free and localized excitons becomes ineffective. We then introduced substitutional carbon defects at chalcogen sites (S and Se), which create localized occupied and unoccupied defect states. These states form a defect-induced two-level system capable of supporting single-photon emission. Consequently, non-resonant excitation below the bandgap is sufficient to generate single-photon emission. Using spin-polarized DFT with the HSE06 hybrid functional, we systematically investigated the ground-state, excited-state, and optical properties of these defects in WS$_2$, WSe$_2$, MoS$_2$, and MoSe$_2$ bilayers. \\
\indent Our analysis revealed that both neutral and singly negative charge states are thermodynamically stable. Neutral defects favor the singlet configuration, with ZPLs at the O-band and C-band telecommunication wavelength range. Importantly, we find that the defect position has only a minor influence on both the ZPL and the PL lineshape. In contrast, the computed radiative lifetimes and dipole orientations vary significantly across different systems, providing distinct fingerprints that may facilitate experimental identification of the defects. We note that a prior work on a neutral C$_\text{S}$ defect in WS$_2$ monolayer has shown the existence of a metastable triplet state with potential for spin initialization and readout \cite{10.1038/s41467-022-28876-7}. Although we find that the C$_\text{S}$ and C$_\text{Se}$ defects in bilayers also inherit the triplet configuration, the detailed intersystem crossing of these spin states and their possible use as qubits are beyond the scope of this work. \\
\indent For singly negative defects, which prefer doublet configurations, the spin-down to spin-down optical channel is identified as the dominant transition due to its favorable deep-level localization. Unlike the neutral counterparts, their ZPLs do not consistently fall within the telecommunication range, but instead lie in the near-infrared range. In addition, their phonon sideband behavior, lifetimes, and dipole orientations exhibit distinct defect- and host-dependent trends. These findings provide well-defined spectroscopic signatures for future experimental validation.\\
\indent In summary, our results demonstrated that substitutional carbon defects in bilayer TMDs constitute a viable pathway toward stable, defect-based quantum emitters operating in the near-infrared and telecommunication regimes. By combining the exciton-suppressing properties of bilayers with the localized defect states of carbon impurities, this work establishes a foundation for room-temperature quantum light sources in TMD materials. 

\section{Method}
\subsection{DFT calculation details}
All first-principles calculations were carried out using the Vienna \textit{Ab initio} Simulation Package (VASP) \cite{vasp1,vasp2}. Pseudopotentials treated by the projector augmented wave (PAW) method are chosen to account for the nucleus and valence electrons \cite{paw,paw2}. To accurately capture the electronic structure of 2H (AA'-stacked) WS$_2$, WSe$_2$, MoS$_2$, and MoSe$_2$ bilayers, we employed the screened hybrid functional HSE06 \cite{10.1063/1.1564060}, which yields band gaps in close agreement with experimental values. Defects were modeled within a $7\times7\times1$ supercell (294 atoms), with a vacuum spacing of 15~\AA\ along the $z$-axis to eliminate spurious interactions between periodic images. This supercell size was confirmed to be sufficient to minimize finite-size effects (see Supplementary S1).  Brillouin-zone integrations were performed using a single $\Gamma$-point scheme, with a plane-wave cutoff energy of 500 eV. Spin polarization was included in all calculations to properly account for spin-dependent electronic transitions. Experimentally, while TMD bilayers can adopt several stacking arrangements; here, we considered only the 2H (AA') configuration, where transition-metal atoms in one layer sit above chalcogen atoms in the adjacent layer. Long-range van der Waals interactions were included using the D3 correction of Grimme \textit{et al.} \cite{10.1063/1.3382344}. For the lattice structural optimization, only internal coordinates were allowed to relax until the residual forces on all atoms were less than 0.01 eV$\cdot$\text{\AA}$^{-1}$ and the total energy was converged to within $10^{-4}$ eV.

\subsection{Defect formation energy}
The stability of each defect and its possible charge states was assessed by calculating the formation energy according to the standard expression
\begin{eqnarray}
    E^{f}[C^{q}] &=& E_{\text{tot}}[C^{q}] - E_{\text{tot}}[\text{TMD}] - \sum_{i}n_i \mu_i \nonumber \\
    &+& q(\epsilon_{\text{vbm}} + \epsilon_{\text{Fermi}}) + E_{corr}(q),
\end{eqnarray}
where $E_{\text{tot}}[C^{q}]$ is the total energy of a supercell containing a carbon defect in charge state $q$; and $E_{\text{tot}}[\text{TMD}]$ is the total energy of the pristine TMD host. The term $n_i$ denotes the number of atoms added or removed, with $\mu_i$ being the chemical potential of the species $i$. The Fermi level $\epsilon_{\text{Fermi}}$ is defined relative to the valence band maximum (VBM) of the pristine TMD, $\epsilon_{\text{vbm}}$. The final term, $E_{corr}(q)$, corrects for spurious electrostatic interactions between charged defects and their periodic images \cite{10.1103/PhysRevLett.102.016402}. Charge corrections were performed using the Spinney Python package \cite{10.1016/j.cpc.2021.107946}.  \\
\indent By systematically evaluating $E^{f}[C^{q}]$ across all charge states, we identify the thermodynamically most stable configuration for each defect. These stable charge states will be explored for the subsequent ground- and excited-state property analyses.

\subsection{Configuration coordinates}
The structural differences between the ground and excited states are quantified using the configuration coordinate ($q_k$), defined as
\begin{equation}
    q_k = \sum_{\alpha,i \in \{x,y,z\}} \sqrt{m_\alpha}\left(R_{e,\alpha i} - R_{g,\alpha i}\right)\Delta r_{k,\alpha i},
\end{equation}
where $\alpha$ labels the atoms; $i$ denotes the Cartesian directions $(x, y, z)$; and $m_{\alpha}$ is the atomic mass of atom $\alpha$. $R_{g,\alpha i}$ and $R_{e,\alpha i}$ represent the equilibrium positions of atom $\alpha$ in the ground and excited states, respectively, while $\Delta r_{k,\alpha i}$ is the normalized eigenvector of phonon mode $k$, specifying the atomic displacement direction.

\subsection{Photoluminescence Spectrum} \label{sec:PL}
The PL spectrum, $L(\hbar\omega)$, is obtained using a full-phonon approach \cite{10.1088/1367-2630/16/7/073026} based on the PyPhotonics workflow \cite{10.1016/j.cpc.2021.108222} and Phonopy \cite{phonopy-phono3py-JPCM,phonopy-phono3py-JPSJ} employed to generate the phonon modes. For the sake of completeness, the principle is summarized as follows. The emission spectrum is expressed as
\begin{equation}
    L(\hbar\omega) = C \, \omega^3 \, A(\hbar\omega),
\end{equation}
where $C$ is a normalization constant, typically adjusted to match experimental intensities, and $A(\hbar\omega)$ is the optical spectral function including the ZPL and its phonon sidebands. The spectral function is computed through
\begin{equation}
    A(E_{ZPL} - \hbar\omega) = \frac{1}{2\pi}\int_{-\infty}^{\infty} G(t) \, e^{-i\omega t - \gamma |t|} \, dt, 
\end{equation}
where $\gamma$ accounts for spectral broadening and $G(t)$ is the generating function defined as
\begin{equation}
    G(t) = \exp[S(t) - S(0)].
\end{equation}
The time-dependent function $S(t)$ is obtained from the phonon spectral distribution,
\begin{equation}
    S(t) = \int_0^\infty S(\hbar\omega) \, e^{-i\omega t} \, d(\hbar\omega),
\end{equation}
where $S(\hbar\omega)$ represents the HR spectral function. The HR factor for each individual phonon mode $k$ is defined as
\begin{equation}
    s_k = \frac{\omega_k q_k^2}{2\hbar},
\end{equation}
with $\omega_k$ being the phonon frequency and $q_k$ the corresponding configuration coordinate displacement. The total HR factor is then obtained by summing over all modes, leading to the spectral representation
\begin{equation}
    S(\hbar\omega) = \sum_k s_k \, \delta(\hbar\omega - \hbar\omega_k),
\end{equation}
where $\delta$ is the Dirac delta function.  \\
\indent Additionally, the DW factor is evaluated to quantify the fraction of emission occurring in the ZPL, given by
\begin{equation}
    \text{DW} = e^{-S}.
\end{equation}
We note that the convergence test of the supercell size was tested, as shown in Supplementary S1.

\subsection{Radiative Lifetime}
The radiative lifetime in this study is determined as the inverse of the radiative transition rate, which is expressed as
\begin{equation}
\Gamma_{\mathrm{R}} = \frac{n_D e^2}{3 \pi \epsilon_0 \hbar^4 c^3} E_0^3 \, \mu_{\mathrm{e}-\mathrm{h}}^2,
\end{equation}
where $\Gamma_{\mathrm{R}}$ represents the radiative transition rate; $e$ is the elementary charge; $\epsilon_0$ is the vacuum permittivity; $E_0$ is the ZPL energy; and $\mu_{\mathrm{e}-\mathrm{h}}^2$ is the squared magnitude of the transition dipole moment, computed as in Eq.~\eqref{eq:dipole}. The term $n_D$ denotes the refractive index of the host bilayer, with values of 4.3751 for WS$_2$, 5.1319 for WSe$_2$, 4.3595 for MoS$_2$, and 4.97 for MoSe$_2$ \cite{10.1002/adom.201900239,10.1038/s41597-023-02898-2}.

\subsection{Transition Dipole Moment}
The transition dipole moment $\boldsymbol{\mu}$ is evaluated using the equilibrium wavefunctions of the ground and excited states, and is given by
\begin{equation}
\boldsymbol{\mu} = \frac{i\hbar}{(E_{f} - E_{i}) m} \bra{\psi_f} \mathbf{p} \ket{\psi_i},
\label{eq:dipole}
\end{equation}
where $E_{i}$ and $E_{f}$ are the eigenvalues of the initial and final orbitals, respectively; $m$ is the electron mass; and $\mathbf{p}$ is the momentum operator. The wavefunctions $\psi_i$ and $\psi_f$ correspond to the equilibrium states of the ground and excited configurations. Since these wavefunctions are generally not identical, they are extracted independently using a modified version of the PyVaspwfc code \cite{pyWave,
10.1088/1361-648X/ab94f4}.  

To resolve the directional components of the dipole, the magnitude of the dipole vector, $\boldsymbol{\mu} = (\mu_x, \mu_y, \mu_z)$, is calculated as
\begin{equation}
r = \sqrt{\mu_x^2 + \mu_y^2 + \mu_z^2}.
\end{equation}
To describe the dipole orientation, the Cartesian components are converted to spherical coordinates. The polar angle $\theta$ is measured from the $z$-axis, while the azimuthal angle $\phi$ is defined in the $xy$-plane from the $x$-axis, expressed as
\begin{equation}
\theta = \arccos\left(\frac{\mu_z}{r}\right), \quad
\phi = \arctan(\frac{\mu_y}{ \mu_x}).
\end{equation}

\section*{Data availability}
All raw data from this work is available from the authors upon reasonable request.

\section*{Notes}
The authors declare no competing financial interest.

\begin{acknowledgments}
This research is part of the Munich Quantum Valley, which is supported by the Bavarian state government with funds from the Hightech Agenda Bayern Plus. This work was funded by the Deutsche Forschungsgemeinschaft (DFG, German Research Foundation) under Germany's Excellence Strategy- EXC-2111-390814868 (MCQST) and as part of the CRC 1375 NOA project C2 (Projektnummer 398816777). The authors acknowledge support from the Federal Ministry of Research, Technology and Space (BMFTR) under grant number 13N16292 (ATOMIQS). S.S. acknowledges research funding by Mahidol University (Fundamental Fund FF-111/2568: fiscal year 2025 by the National Science Research and Innovation Fund (NSRF)). The authors gratefully acknowledge the Gauss Centre for Supercomputing e.V.\ (www.gauss-centre.eu) for funding this project by providing computing time on the GCS Supercomputer SuperMUC-NG at Leibniz Supercomputing Centre (www.lrz.de) and on its Linux-Cluster.
\end{acknowledgments}

\section*{Author contributions}
T.V. conceived the project. C.C. performed and analyzed the calculations. All authors contributed to the discussion and review of the manuscript. T.V. and S.S. supervised the project.

\bibliography{main}
\end{document}